\documentclass[aps,prd,nofootinbib,10pt,tightenlines,noti, twocolumn,floatfix,superscriptaddress,showkeys]{revtex4-2}

\usepackage{makecell}

\usepackage[utf8]{inputenc}          
\usepackage{graphicx}         
\usepackage[usenames,dvipsnames]{xcolor}
\usepackage{array,dcolumn,longtable} 
\usepackage{amsmath,amssymb,amsfonts,slashed} 
\usepackage{mathtools}          
\usepackage[linktocpage,breaklinks]{hyperref}
\usepackage{mathrsfs}
\usepackage{txfonts}
\usepackage{bm}
\usepackage{stmaryrd}
\usepackage{tensor}
\usepackage[utf8]{inputenc}

\usepackage{epsfig}
\usepackage{epstopdf}

\usepackage{cleveref}

\definecolor{mred}{RGB}{127,0,25}
\definecolor{mdgr}{RGB}{51,51,51}
\definecolor{mag}{RGB}{211, 54, 130}
\definecolor{verm}{RGB}{164, 25, 0}

\hypersetup{colorlinks=true,
            citecolor=NavyBlue,
            linkcolor=NavyBlue,
            urlcolor=NavyBlue}
\usepackage{siunitx}                                  
\DeclareSIUnit{\fm}{\femto\metre}                     

\newcount\hh
\newcount\mm
\mm=\time
\hh=\time
\divide\hh by 60
\divide\mm by 60
\multiply\mm by 60
\mm=-\mm
\advance\mm by \time
\def\hhmm{\number\hh:\ifnum\mm<10{}0\fi\number\mm}

\def\be{\begin{equation}}
\def\ee{\end{equation}}

\def\ve{\varepsilon}
\def\tH{{\tilde H}}
\def\tom{{\tilde \Omega}}

\begin{document}

\preprint{APS/123-QED}

\title{Motion of a spinning particle under the conservative piece\\ of the self-force is Hamiltonian to first order in mass and spin}

\author{Francisco M. Blanco}
\affiliation{Department of Physics, Cornell University, Ithaca, NY 14853, USA.}

\author{Éanna É. Flanagan}
\affiliation{Department of Physics, Cornell University, Ithaca, NY 14853, USA.}


\begin{abstract}

We consider the motion of a point particle with spin in a stationary
spacetime. We define, following Witzany \cite{witzanyham} and later Ramond  \cite{paulramond}, a twelve
dimensional Hamiltonian dynamical system whose orbits coincide with
the solutions of the Mathisson-Papapetrou-Dixon equations of motion with
the Tulczyjew-Dixon spin supplementary condition, to linear order in
spin. We then perturb this system by adding the conservative pieces of
the leading order gravitational self-force and self-torque sourced by
the particle's mass and spin. We show that this perturbed system is
Hamiltonian and derive expressions for the Hamiltonian function and
symplectic form. This result extends a previous result for spinless
point particles \cite{Blanco:2022mgd}.

\end{abstract}

\maketitle

\vspace{0.1cm}

\begin{section}{\label{sec:intro} Introduction}

In recent years, the detection of coalescences of binary black hole
systems has started a new era of gravitational wave astronomy
\cite{introLIGO1,introLIGO2,introLIGO3}.
The coming years will bring many more detections with 
the next generation ground based detectors Cosmic Explorer
\cite{Evans:2021gyd} and the Einstein Telescope
\cite{2010CQGra..27s4002P}, the space based detector LISA
\cite{Audley:2017drz}, and potentially pulsar timing arrays
\cite{Antoniadis:2022pcn}. The observation of gravitational waves
requires precise waveform templates, which for binary coalescences can be obtained through
a variety of different approximation methods valid in different regimes. Some of the
techniques that have been used to understand the dynamics of black
hole binaries are numerical relativity \cite{Lehner:2014asa}, the  
post-Newtonian approximation
\cite{introPN,poissonwill,Levi:2018nxp,Porto:2016pyg}, the
post-Minkowskian approximation \cite{Damour:2016gwp} for which
amplitude methods from quantum field theory are
useful \cite{Bern:2021dqo}, the small mass ratio approximation
\cite{introEMRI,pound}, 
and the effective one-body framework which synthesizes information
from the other approaches \cite{Damour:2012mv,Taracchini:2013rva}.

A theoretical issue that arises in the study of binary dynamics is
whether or not the motion forms a Hamiltonian dynamical system when
gravitational wave dissipation is turned off. This has been
established to various orders in the post-Newtonian and
post-Minkowskian approximations  (see Ref.\ \cite{hamiltonians} and references
therein). For non-spinning particles, it has also been established to
first order in the small mass ratio approximation
\cite{Blanco:2022mgd}. The small mass-ratio approximation consists of
an expansion in the ratio $\epsilon  = \mu/M$ of the mass $\mu$ of
the secondary object to the mass $M$ of the primary object. The
gravitational field of the secondary acts as a perturbation to the
background  geometry, which can be expanded in powers of
$\epsilon$. The interaction between the secondary and its own
gravitational field gives rise to an acceleration with respect to the
background geometry, described by the gravitational self-force
\cite{misata,quwa}. The self-force itself can be divided into
conservative and dissipative pieces. The former is derived from the
time symmetric piece of the Green's function while the latter comes
from its time antisymmetric piece and is responsible for the
dissipation that drives the slow inspiral. In previous
work \cite{Blanco:2022mgd}, we showed that the conservative piece of
the first order self-force gives rise to Hamiltonian dynamics, and derived
an explicit expression for the Hamiltonian. The goal of this
paper is to extend that result to include the leading spin effects of
the secondary. 

The motion of a point particle with spin in general relativity,
neglecting self gravity, is described by the
Mathisson-Papapetrou-Dixon equations \cite{mpdm,mpdp,mpdd}.
A variety of Hamiltonian formulations of the dynamics in the test body
limit have been given in \cite{ham1,ham2,witzanyham,paulramond}.
Many of these formulate the dynamics as a constrained Hamiltonian system.  We will
follow an approach developed by Witzany \cite{witzanyham} and later
generalized by Ramond \cite{paulramond} which yields  
an unconstrained Hamiltonian system on a twelve dimensional space.
Going beyond the test body limit to include self gravity and working
to leading order in spin, the motion is described by a first
order self-force which depends on mass and spin, and by a first order
self-torque \cite{Mathews:2021rod}.  Specifically the
self-force has terms of order $O(\mu^2)$, $O(S)$ and $O(\mu S)$, where
$\mu$ is mass and $S$ spin, and the self-torque scales as $O(\mu S)$.
We will show that this dynamical system is also Hamiltonian, and will
derive the explicit form of the Hamiltonian.

\begin{table*}[t] \label{table}
\begin{tabular}{|c|c|c|c|c|c|}
\hline
\thead{Name of effect} & \thead{Interaction \\ energy} & \thead{Accumulated phase\\shift in inspiral} & \thead{Papers that \\ discuss effect}& \thead{Previously known\\ Hamiltonian?} & \thead{Included \\ in this paper?} \\
\hline  
\hline
Geodesic motion&$\sim \mu$&\makecell{$\sim \frac{1}{\epsilon}$}&& $\checkmark$ &$\checkmark$ \\
\hline
First order conservative self-force&$\sim \mu \epsilon$&$\sim 1$&\cite{tanaka}&$\checkmark$ \cite{Blanco:2022mgd}&$\checkmark$ \\
\hline
Second order conservative self-force&$\sim \mu \epsilon ^2$&$\sim \epsilon$&\cite{Pound:2014koa,Pound:2019lzj,gralla,pound1,Bini:2016cje}&&\\ 
\hline
Leading spin-curvature coupling&\makecell{$\sim \frac {S}{M}\sim \mu
  \epsilon$}&$\sim \frac{S}{\mu^2} \sim 1$&\cite{semerak1,semerak2,mpdm,mpdp,mpdd,Drummond:2022efc,Drummond:2022xej,ham1}&$\checkmark$ \cite{witzanyham,paulramond}&$\checkmark$\\
\hline
Subleading spin-curvature coupling&\makecell{$\sim \frac{S^2}{\mu
    M^2}\sim \mu \epsilon^2$}&$\sim \frac{S^2}{\mu^3 M} \sim \epsilon$&\cite{ham2}&$\checkmark$ \cite{ham2}&\\
\hline
\makecell{First order conservative spin-induced\\ self-force and
  self-torque}&\makecell{$\sim \frac{S\mu}{M^2}\sim \mu
  \epsilon^2$}&$\sim \frac{S}{\mu M} \sim \epsilon$&\cite{Mathews:2021rod,Drummond:2022efc,Drummond:2022xej}&&$\checkmark$\\
\hline
\end{tabular}
\caption{A summary of different conservative effects in the dynamics
  of binary systems with spinning components in general relativity, in
  the small mass ratio limit $\mu \ll M$, where $M$ is the mass of the
  primary and $\mu$ the mass of the secondary.  The effects can be
  distinguished by the scaling of the associated interaction energy
  with the parameters of the system. The first column lists the
  effects, and the second the interaction energies, where $\epsilon =
  \mu/M$ is the mass ratio and $S$ is the spin of the secondary.
  The third column lists the total phase shift caused by the effect
  that is accumulated during the dissipative self-force driven
  inspiral through the relativistic 
  regime.
  The second entry in each box is the result specialized to an order unity
  dimensionless spin parameter $S/\mu^2 \sim 1$ for the secondary.
  Some previous discussions of these effects in the literature are
  listed in the fourth column.  The fifth column specifies if a
  Hamiltonian formulation of the effect was previously known and gives
  references.  Finally the last column indicates which effects are
  included in the analysis of this paper.}
\end{table*}

The various spin-related conservative effects that arise in the
dynamics of two body systems are listed in Table I, which includes the
scaling of the interaction energy and accumulated orbital phase with
the parameters of the system.  Effects which were previously known to
admit Hamiltonian formulations include geodesic motion,
the first order point-particle self-force \cite{Blanco:2022mgd}, and
the leading \cite{witzanyham,paulramond} and
subleading \cite{ham2} spin-curvature couplings.  The effect for which we give a
new Hamiltonian formulation is the self-interaction
associated with spin, and is listed in the last row.
It consists of two different pieces which enter at the same order
\cite{Mathews:2021rod}.  First, the regularized
self-field of the secondary has a contribution of order $\sim \mu/M$, and the
spin-curvature coupling force as well as the spin parallel transport get corrections due this metric perturbation.
Second, there is a contribution to the regularized self-field of order $\sim
S/M^2$, and the gradient of this field gives a correction to the self-force.

For extreme mass ratio inspirals with near maximal spin $S \sim
\mu^2$, the accumulated phase shift over an inspiral due to the spin
related self-interaction scales as $\mu/M \ll 1$, as shown in
Table \ref{table}.  Hence the Hamiltonian we derive will not be
relevant for computations of waveforms for LISA, for which
post-1-adiabatic waveforms which include all the $O(1)$ effects in the
accumulated phase will suffice.  However, it may yield
useful information for the effective one body framework \cite{Damour:2012mv,Taracchini:2013rva}
and thereby
aid waveform modeling for comparable mass binary systems for LIGO.
We also note that our analysis does not include second order
point-particle self-force effects, and subleading spin-curvature
effects, 
even though they enter at the same order as the spin self-interaction effect when $S \sim
\mu^2$ as shown in Table \ref{table}.  It would be interesting to extend our analysis to include
these effects.

The organization of this paper is as follows.  In
Sec.\ \ref{sec:testham} we review the dynamics of a test spinning
particle up to linear order in spin, given by the
Mathisson-Papapetrou-Dixon equations. We specialize to the 
Tulczyjew-Dixon spin supplementary condition and review the
Hamiltonian formulation of the resulting dynamical system.
The existence of two Casimir invariants makes the Poisson
brackets degenerate.  By passing to the submanifold of the phase space
on which the Casimirs are constant, we obtain a true Hamiltonian
dynamical system with nondegenerate Poisson brackets, following
Witzany \cite{witzanyham} and Ramond \cite{paulramond}.
In Sec.\ \ref{sec:pseudoham} we define pseudo-Hamiltonian dynamical systems and review
a general result in the theory of these systems that gives sufficient
conditions for a pseudo-Hamiltonian system to be Hamiltonian \cite{Blanco:2022mgd}.
We derive in Sec.\ \ref{sec:pseudohamforslefforce} a
pseudo-Hamiltonian
formulation of the dynamics of a spinning point
particle including self-force effects.
This is obtained by replacing the metric in the test-particle
Hamiltonian by an effective metric, which includes perturbations proportional to the particle's
mass and spin. Lastly,
in Sec.\ \ref{sec:ham} we apply the result from section
\ref{sec:pseudoham} to obtain a Hamiltonian description of the motion
of a spinning particle.

Throughout this paper we use geometric units with $G = c = 1$.

\end{section}


\begin{section}{\label{sec:testham} Hamiltonian description of the motion of a spinning test particle}

The motion of an extended body in general relativity, neglecting self gravity, can be reduced to the motion of a point particle of mass $\mu$ endowed with a series of mass and current multipole moments \cite{dixon1,tulcz1,lukes1,Harte_2015}. If we restrict ourselves to the pole-dipole approximation, where only the
mass and spin are included, the dynamics are given by the well-known Mathisson-Papapetrou-Dixon (MPD) equations \cite{mpdm,mpdp,mpdd}  
\begin{subequations} \label{eq:MPD1}
  \begin{eqnarray}
\nabla_{\vec{u}}p_\mu&=&-\frac{1}{2}R_{\mu\nu\alpha\beta}u^\nu S^{\alpha\beta},\\
\nabla_{\vec{u}}S^{\alpha\beta}&=& 2p^{[\alpha}u^{\beta]}    .
  \end{eqnarray}
\end{subequations}
Here
\begin{eqnarray} \label{eq:MPD2}
  \frac{dx^\mu}{d\tau} = u^\mu
\end{eqnarray}
is the 4-velocity of the particle, $S^{\alpha\beta}$ is its spin tensor, 
$p_\mu$ is its 4-momentum, $\nabla_{\vec{u}}=u^\alpha \nabla_\alpha$
is the covariant derivative respect to proper time $\tau$, and $R_{\mu \nu \alpha \beta}$ is the Riemann tensor.
The set of equations (\ref{eq:MPD1}) and (\ref{eq:MPD2}) comprises 14 equations for 17 independent unknowns $x^\mu(\tau)$, $u^\mu(\tau)$, $p_\mu(\tau)$ and $S^{\alpha\beta}(\tau)$. Hence the dynamical system is not yet completely specified. This incompleteness arises because of the freedom to choose different definitions of the center-of-mass worldline $x^\mu(\tau)$ of the extended body \cite{semerak1,semerak2}. A definition can be chosen by imposing a so-called spin supplementary condition of the form
\begin{equation}
    S^{\alpha\beta}V_\beta =0,
\end{equation}
for some timelike vector $V_\beta$. 

In this paper, we use the Tulczyjew-Dixon spin supplementary condition \cite{tulcz1,mpdd}
\be
S^{\alpha\beta}p_\beta=0,
\label{TD}
\ee
which reduces the MPD
equations to \cite{Drummond:2022xej}
\begin{subequations}\label{eq:prempd}
\begin{eqnarray}
 \label{prempd1} \frac{dx^\mu}{d\tau}&=&\frac{1}{\mu}g^{\mu\nu}p_\nu-\frac{1}{2\mu^3}R_{\sigma\rho\alpha\beta}S^{\sigma\mu}p^\rho S^{\alpha\beta}+O(S^4),\\
\label{prempd2}  \nabla _{\vec{u}}p_\mu&=&-\frac{1}{2\mu}R_{\mu\nu\alpha\beta}S^{\alpha\beta}p^\nu+O(S^3),\\
\label{prempd3}\nabla_{\vec{u}}S^{\alpha\beta}&=&-\frac{1}{\mu}u^{[\alpha}S^{\beta]\sigma}u^\rho S^{\mu\nu}R_{\sigma\rho\mu\nu}+O(S^4).
\end{eqnarray}
\end{subequations}
Here we have dropped terms cubic and higher order in spin and have defined the particle mass
\be
\label{eq:mpd0} \mu=\sqrt{-g^{\alpha\beta}p_\alpha p_\beta}.
\ee

The dynamics to leading order in spin is obtained by 
dropping all of the terms quadratic in spin in
Eqs.\ (\ref{eq:prempd}), and keeping only the term in the momentum
evolution equation (\ref{prempd2}) which is linear in spin
\cite{Drummond:2022xej,paulramond}.
This term corresponds to the leading spin-curvature coupling effect
listed in Table \ref{table}.  In this approximation 
the MPD equations (\ref{eq:prempd}) reduce to
\begin{subequations}\label{eq:mpd}
\begin{eqnarray}
\label{eq:mpd1}  \frac{dx^\mu}{d\tau}&=&\frac{1}{\mu}g^{\mu\nu}p_\nu\\
\label{eq:mpd2}  \nabla _{\vec{u}}p_\mu&=&-\frac{1}{2\mu}R_{\mu\nu\alpha\beta}S^{\alpha\beta}p^\nu\\
\label{eq:mpd3}\nabla_{\vec{u}}S^{\alpha\beta}&=&0.
\end{eqnarray}
\end{subequations}

We note that the spin supplementary
condition (\ref{TD}) is not preserved by the dynamics (\ref{eq:mpd}).  
This arises because we are working to linear order in spin.  In this
paper we shall adopt the equations (\ref{eq:mpd}) as the definition of the
dynamical system we are working with, even though this definition is
formally inconsistent with the spin supplementary condition from which
it was derived.  The inconsistency is higher order in spin and so can
be safely ignored for our purposes.

Let $\Gamma_s$ denote the phase space consisting of the bundle over
spacetime with coordinates $(x^\mu,p_\nu,S^{\alpha\beta})$. As is well
known, there exists a Hamiltonian function and a Poisson bracket
structure on $\Gamma_s$ that give rise to the dynamical system
(\ref{eq:mpd})
\cite{ham1,ham2,witzanyham,paulramond,van_Holten_2016,d_Ambrosi_2015}. The
Poisson brackets are
\begin{subequations}\label{eq:poisson}
\begin{eqnarray}
    \{x^\mu,x^\nu\}&=&0,\\
    \{x^\mu,p_\nu\}&=&\delta^\mu _\nu , \\
    \{p_\mu,p_\nu\}&=&-\frac{1}{2}R_{\mu\nu\alpha\beta}S^{\alpha\beta},\\
    \{x^\mu,S^{\alpha\beta}\}&=&0,\\
    \{S^{\alpha\beta},p_\mu\}&=&-\Gamma^\alpha_{\mu\rho}S^{\rho\beta}-\Gamma^\beta_{\mu\rho}S^{\alpha\rho},\\
    \{S^{\mu\nu},S^{\alpha\beta}\}&=&2g^{\mu[\beta}S^{\alpha]\nu}-2g^{\nu[\beta}S^{\alpha]\mu},
\end{eqnarray}
\end{subequations}
and the Hamiltonian $H_0$ is
\be \label{eq:geoham}
H_0(x,p,S)=-\sqrt{-g^{\mu \nu}p_{\mu}p_{\nu}}.
\ee

It will be convenient to make a change of coordinates on phase space
to simplify the form (\ref{eq:poisson}) of the Poisson brackets \cite{quasican}. We
choose an arbitrary orthonormal basis $\vec{e}_\Lambda=e_\Lambda
^{\ \alpha} \partial _\alpha$ for $0\leq\Lambda \leq 3$, with
$\vec{e}_\Lambda \cdot \vec{e}_\Sigma = \eta_{\Lambda \Sigma}$, the
Minkowski metric with signature $(-1,1,1,1)$. We use upper case Greek
indices for orthonormal basis indices and lower case Greek indices for
spacetime indices. We define the dual basis ${\bf e}^\Lambda =
e^\Lambda_{\ \mu} dx^\mu$ by $e^\Lambda_{\ \mu} e^{\ \mu}_\Sigma =
\delta^\Lambda_\Sigma$, and the components of the spin connection by  
\begin{equation}
    \omega_{\alpha \Lambda \Sigma}= e_{\Lambda \rho}\nabla_\alpha e_\Sigma^{\ \rho}.    
\end{equation}
We define new phase space coordinates $(x^\alpha,\pi_\alpha,S^{\Lambda\Pi})$ by
\begin{subequations}\label{eq:quasisym}
  \begin{eqnarray}
    \label{pidef}
        \pi_\alpha&=& p_\alpha - \frac{1}{2}\omega_{\alpha \Lambda
          \Sigma} \, e^\Lambda_{\ \mu} \, e^\Sigma_{\ \nu} \, S^{\mu \nu}, \\
        S^{\Lambda\Sigma}&=& e^\Lambda_{\ \mu} \, e^\Sigma_{\ \nu}\, S^{\mu \nu}.
    \end{eqnarray}
\end{subequations}
In these new coordinates the only non-vanishing Poisson brackets are
\begin{subequations} \label{eq:poisson1}
\begin{eqnarray}
\{x^\mu,\pi_\nu\}&=&\delta ^\mu _\nu, \\
\{S^{\Theta \Pi},S^{\Gamma \Lambda}\}&=&2\eta^{\Theta[\Lambda}S^{\Gamma]\Pi}-2\eta^{\Pi[\Lambda}S^{\Gamma]\Theta}    .
\end{eqnarray}
\end{subequations}
Substituting the coordinate change (\ref{eq:quasisym}) into the Hamiltonian (\ref{eq:geoham}) and linearizing in spin gives the form of the Hamiltonian in these coordinates
\be
\label{eq:mpdham}
H_0(x,\pi,S)=-\sqrt{-g^{\mu\nu}\pi_\mu \pi_\nu}+\frac{g^{\mu\nu}\pi_\mu \omega_{\nu \Theta \Pi}S^{\Theta \Pi}}{2\sqrt{-g^{\mu\nu}\pi_\mu \pi_\nu}}.
\ee
It will also be convenient to define a new mass parameter $m$ related to
the norm of the new momentum 4-vector
\begin{equation}
    m=\sqrt{-g^{\alpha\beta}\pi_\alpha \pi_\beta},
\end{equation}
which is related to our previously defined mass (\ref{eq:mpd0}) by
$m=\mu+O(S)$. In the following sections we will expand the
Hamiltonian of the system in powers of $m$ and $S$, by counting
factors of $\pi_\mu$ and $S^{\Lambda \Pi}$. Using this counting the first
term in the Hamiltonian (\ref{eq:mpdham}) is $O(m)$ while the second one is
$O(S)$.

Although the Hamiltonian function (\ref{eq:mpdham}) and Poisson
structure (\ref{eq:poisson1}) give rise to the dynamical system
(\ref{eq:mpd}) on $\Gamma_s$, the dynamical system is not Hamiltonian 
since the Poisson structure (\ref{eq:poisson1}) is degenerate.
The degeneracy is
due to the existence of two Casimir invariants\footnote{The quantities $S_{*}^2$ and $S_{\circ}^2$ are intended to be interpreted
the same way as the square of a four-momentum, that is, they can have either sign despite the notation as a square.} \cite{witzanyham,paulramond}
\begin{subequations} \label{eq:casimirs}
\begin{eqnarray}
S_{*}^2&=&\frac{1}{8}\epsilon_{\Gamma\Sigma\Xi\Pi}S^{\Gamma\Sigma}S^{\Xi\Pi}, \\
S_{\circ}^2&=&\frac{1}{2}\eta_{\Gamma\Sigma}\eta_{\Xi\Pi}S^{\Gamma\Xi}S^{\Sigma\Pi},
\end{eqnarray}
\end{subequations}
which satisfy
\be
\{S_*^2,F\}=\{S_\circ^2,F\}=0
\label{casimireqn}
\ee
for any function $F$ on phase
space.
Denoting by $y^{\cal A}$ abstract coordinates on $\Gamma_s$, the Poisson
structure can be written as a tensor $\Omega^{{\cal A}{\cal B}}$, and
its degeneracy implies that a symplectic form $\Omega_{{\cal A}{\cal B}}$
satisfying
$\Omega_{{\cal A}{\cal B}} \Omega^{{\cal B}{\cal C}} = \delta^{\cal
  A}_{\cal C}$
does not exist. Thus, $\Gamma_s$ is
a Poisson manifold but not a symplectic manifold.

We can overcome this difficulty and obtain a true Hamiltonian
description of the dynamics as follows, following \cite{witzanyham,paulramond}.  Fix values $S_\circ^2$ and $S_*^2$ of the Casimirs,
and consider the corresponding submanifold $\Gamma$ of $\Gamma_s$.
Denoting by $Q^A$ abstract coordinates on $\Gamma$, and by $y^{\cal A}
= y^{\cal A}(Q^B)$ the embedding map. There 
exists an invertible Poisson structure $\Omega^{AB}$ on $\Gamma$ whose
pushforward
\be
\Omega^{{\cal A}{\cal B}} = \frac{\partial y^{\cal A}}{\partial Q^A}
\frac{\partial y^{\cal B}}{\partial Q^B} \Omega^{AB}
\ee
to $\Gamma_s$ coincides with the Poisson structure
(\ref{eq:poisson1}). It follows that the
dynamical vector field $v^{\cal A} = \Omega^{{\cal A}{\cal B}}
\partial_{\cal B} H_0$ on $\Gamma_s$ is the pushforward $v^A \partial
y^{\cal A} / \partial Q^A $ of the Hamiltonian vector field $v^A =
\Omega^{AB} \partial_B {\bar H}_0$ on $\Gamma$, where ${\bar H}_0$ is
the pullback of $H_0$ to $\Gamma$ (below we will drop the bar).  Thus, the dynamics restricted to
$\Gamma$ is Hamiltonian and $\Gamma$ is a symplectic manifold.

We will restrict attention to submanifolds $\Gamma$ of $\Gamma_s$ for which
\be
S_\circ ^2\geq0, \ \ \ \ S_*^2= 0.
\label{ccc}
\ee
The conditions (\ref{ccc}) will be satisfied at some point along the trajectory if 
there exists a timelike vector
$f_\beta$ for which
\be
S^{\alpha\beta}f_\beta=0
\label{ccc1}
\ee
at that point \cite{paulramond}.  Then the conditions (\ref{ccc}) will be satisfied at all points along the trajectory, since the quantities $S_\circ^2$ and $S_*^2$ are conserved by the dynamics
by Eq.\ (\ref{casimireqn}).
Our spin supplementary condition (\ref{TD}) satisfies the criterion (\ref{ccc1}).  Although this condition is not preserved by the dynamical evolution (\ref{eq:mpd}), as noted above, if we choose initial data that satisfy the spin supplementary condition then the conditions (\ref{ccc}) will be satisfied all along the trajectory. 
Thus, without loss of any physical generality, we can restrict
attention to submanifolds $\Gamma$ satisfying (\ref{ccc})
\cite{witzanyham,paulramond}.

We now review the construction of the nondegenerate Poisson structure $\Omega^{AB}$
on $\Gamma$ \cite{paulramond}.  We define coordinates
$\{x^\mu,\pi_\mu,\sigma,\rho_\sigma,\varsigma,\rho_\varsigma\}$ on $\Gamma$ by the relations
\begin{subequations} \label{eq:canonicalangles}
\begin{eqnarray}
S^{23}&=&X \cos \sigma, \\
S^{31}&=& X \sin \sigma,\\ 
S^{12}&=& \rho_\sigma,\\ 
S^{01}&=& Y\rho_\sigma \sin \varsigma \cos \sigma + Y \rho_\varsigma \cos \varsigma \sin \sigma + XZ \cos \sigma,  \\ 
S^{02}&=& Y \rho_\sigma \sin \varsigma \sin\sigma - Y \rho_\varsigma \cos\varsigma \cos\sigma +XZ \sin \sigma,  \\ 
S^{03}&=& Z \rho_\sigma - XY \sin \varsigma,
\end{eqnarray}
\end{subequations}
with
\begin{equation}\label{eq:canonicalangles2}
    X=\sqrt{\rho_\varsigma ^2 - \rho_\sigma ^2} ,  \ \ \ \ Y=\sqrt{1-\frac{S_\circ^2 }{\rho_\varsigma^2}-\frac{S_*^2}{\rho_\varsigma^4}}, \ \ \ \ Z=\frac{S_*^2}{\rho_\varsigma^2}.
\end{equation}
We define a Poisson structure 
by
\begin{subequations} \label{eq:canpoisson}
    \begin{eqnarray}
        \{\sigma,\rho_\sigma\}&=&1,\\
       \{\varsigma,\rho_\varsigma\}&=&1,\\
       \{x^\mu,\pi_\nu\}&=&\delta^\mu_\nu,
    \end{eqnarray}
\end{subequations}
with all other brackets vanishing. This is equivalent to the
symplectic form $\mathbf{\Omega}=d\rho_\sigma \wedge d\sigma +
d\rho_\varsigma \wedge d\varsigma + d\pi_\mu \wedge dx^\mu $. One can
check that the pushforward of the Poisson structure (\ref{eq:canpoisson})
using the embedding (\ref{eq:canonicalangles}) and (\ref{eq:canonicalangles2}) gives the Poisson structure (\ref{eq:poisson1}). 

To summarize, the Hamiltonian system on the twelve dimensional phase
space $\Gamma$ is given by the
Poisson brackets (\ref{eq:canpoisson}), and by the Hamiltonian
(\ref{eq:mpdham}) expressed in terms of the coordinates
$\{x^\mu,\pi_\mu,\sigma,\rho_\sigma,\varsigma,\rho_\varsigma\}$ using
the map (\ref{eq:canonicalangles}) and (\ref{eq:canonicalangles2}).

\end{section}


\begin{section}{\label{sec:pseudoham}General result for
    pseudo-Hamiltonian dynamical systems}

In this section we define a class of dynamical systems called
pseudo-Hamiltonian dynamical systems, and review a general result
for these systems \cite{Blanco:2022mgd} which will be the foundation
for the result of this paper derived in Sec. \ref{sec:ham} below. A {\it
  pseudo-Hamiltonian} dynamical system (see \cite{Blanco:2022mgd} for
details) consists of a phase space $\Gamma$, a closed, 
non-degenerate two form $\Omega_{AB}$ and a smooth pseudo-Hamiltonian function
${\cal H} : \Gamma \times \Gamma \to {\bf R}$, for which the dynamics are given by integral curves of the vector field
\be
v^A = \Omega^{AB} \frac{\partial}{\partial Q^B} \left. {\cal H}(Q,Q')
\right|_{Q'=Q}.
\label{phdef}
\ee 

We now specialize to pseudo-Hamiltonian systems which are
perturbations of Hamiltonian systems, with symplectic form
and pseudo-Hamiltonian
\begin{subequations}
  \label{phpfull}
  \begin{eqnarray}
    \label{php00}
  \Omega_{AB} &=& \Omega_{0\,AB},\\
   {\cal H}(Q,Q') &=& H_0(Q) + \ve {\cal H}_1(Q,Q') + O(\ve^2).
\label{php}
\end{eqnarray}
\end{subequations}
Here $\ve$ is a formal expansion parameter. The pseudo-Hamiltonian perturbation ${\cal H}_1$ is defined in terms of a function $G :
\Gamma \times \Gamma \to {\bf R}$ via
\be
\label{php1}
{\cal H}_1(Q,Q') = \int_{-\infty}^\infty d\tau' {\tilde
  G}(0,Q,\tau',Q'),
\ee
where we have defined
\begin{equation}
{\tilde G}(\tau, Q, \tau', Q') = G\left[ \varphi_\tau(Q),
  \varphi_{\tau'}(Q') \right].
\label{tildeGdef}
\end{equation}
Here $\varphi_\tau:\Gamma \rightarrow \Gamma$ is the Hamiltonian
flow associated with the zeroth order Hamiltonian system that takes any point
$\tau$ units along the corresponding integral curve. Writing $Q^A$ for
abstract coordinates on $\Gamma$, the flow satisfies the relations 
\begin{subequations} \label{eq:hamflow}
    \begin{eqnarray}
        \varphi_\tau \circ \varphi_{\tau'}&=&\varphi_{\tau+\tau'}, \\
        \left. \frac{d}{d\tau}\right|_{\tau=0}\varphi^A_\tau (Q)&=&\Omega_0^{AB}\partial_B H_{0} .
    \end{eqnarray}
\end{subequations}
The function $G$ is assumed to 
satisfy the conditions
\begin{subequations}
  \label{condts}
  \begin{eqnarray}
\label{condtA}
G(Q,Q') &=& G(Q',Q), \\
\label{condtB}
{\tilde G} (\tau, Q, \tau',Q') &\to& 0 \ \ {\rm
  as} \ \tau\ {\rm or} \ \tau' \to \pm \infty.
  \end{eqnarray}
  \end{subequations}
  
In Ref.\ \cite{Blanco:2022mgd} we showed that any pseudo-Hamiltonian
dynamical system of the form (\ref{phdef}) and (\ref{phpfull}) can be recast as a
Hamiltonian system, with Hamiltonian and symplectic form
\begin{subequations}
\label{ham11}
  \begin{eqnarray}
\tilde{H}(Q) & =& H_0(Q)+\varepsilon \tilde{H}_1(Q)+O(\varepsilon^2),\\
\tilde{\Omega}_{AB}&=&\Omega^0_{AB}+\varepsilon \tilde{\Omega}^1_{AB}+O(\varepsilon^2).
  \end{eqnarray}
  \end{subequations}
Here the perturbation to the Hamiltonian is
\be
\tilde{H}_1(Q)=\int d\tau' {\tilde G}(0,Q,\tau',Q),
\label{H1def1}
\ee
and the perturbation to the symplectic form is
\be
\tilde{\Omega}^{1}_{AB}(Q)=\left[\frac{\partial}{\partial
    Q^A}\frac{\partial}{\partial Q^{B'}}\int d\tau d\tau'
  \chi(\tau,\tau') {\tilde G}(\tau,Q,\tau',Q')\right]_{Q'=Q}
\ee
where
$\chi(\tau,\tau')=\big[\text{sgn}(\tau)-\text{sgn}(\tau')\big]/2$.

A more convenient representation of the Hamiltonian system
(\ref{ham11}) can be
obtained by performing a linearized diffeomorphism on phase
space \cite{Blanco:2022mgd}.  Under such a diffeomorphism
parameterized by a vector field $\xi^A$, the perturbations to the
Hamiltonian and symplectic form transform as
\begin{subequations}
  \begin{eqnarray}
    \label{newH}
    \tH_1 &\to& H_1 = \tH_1+\mathcal{L}_\xi H_0,\\
    \tom_{1\,AB} &\to& \Omega_{1\,AB} = \tom_{1\,AB} + (\mathcal{L}_\xi \Omega_0)_{AB}.
\end{eqnarray}
\end{subequations}
We choose the linearized diffeomorphism to be 
\be \label{eq:diffeo}
\xi^A = \frac{1}{2} \Omega^{AB}_0  \left[ {\partial \over
    \partial Q^{B'}} \int d\tau \int d \tau'
  \chi\, {\tilde G}(\tau,Q,\tau',Q')
  \right]_{Q'=Q}.
\ee
This yields for the new symplectic form perturbation
\be
\Omega_{1\,AB} =0,
\ee
and the new Hamiltonian
\be
\label{eq:hamresult0}
H(Q) = H_0(Q) + \varepsilon H_1(Q) + O(\varepsilon^2),
\ee
with
\be
\label{eq:hamresult}
H_1(Q)=\frac{1}{2} \int d\tau' {\tilde G}(0,Q,\tau',Q),
\ee
which differs from (\ref{H1def1}) by a factor of $1/2$.

\end{section}

\begin{section}{\label{sec:pseudohamforslefforce} Pseudo-Hamiltonian description of the motion a self-gravitating spinning particle} 

In this section we cast the motion of a spinning particle including
the leading order self-force and self-torque as a pseudo-Hamiltonian
dynamical system of the type discussed in the previous section.  This
will allow us to use the general result discussed there to deduce that
the motion is Hamiltonian.

We start by reviewing the similar
pseudo-Hamiltonian formulation of the motion of a spinless point
particle including the leading order self-force \cite{Blanco:2022mgd}.
For the zeroth order geodesic motion we use phase space coordinates
$Q^A = (x^\mu, p_\mu)$ with symplectic form $\Omega_0 = d p_\mu \wedge dx^\mu$
and Hamiltonian
$H_0 = - \sqrt{ - g^{\mu\nu}(x) p_\mu p_\nu }.
$
For the first order motion, consider a particle at location $x^{\mu'}$
with initial 4-momentum $p_{\mu'}$.  Writing $Q' = (x',p')$, we denote
by
$
\varphi_{\tau'}(Q') = [ x^{{\bar \mu}}(\tau'), p_{{\bar \mu}}(\tau')]
$
the geodesic with initial data $Q'$, where $\tau'$ is proper time.  From this geodesic we can
compute the Lorenz gauge metric perturbation
\be
\label{hpp}
h^{\mu\nu}_R(x,Q') = \frac{1}{\sqrt{ - g^{\mu'\nu'} p_{\mu'} p_{\nu'}}}
\int d\tau'
G_R^{\mu\nu\, {\bar \mu}{\bar \nu}}[x, {\bar x}(\tau')]
p_{{\bar \mu}}(\tau') p_{{\bar \nu}}(\tau').
\ee
Here the symmetric Green's function
$G_R^{\mu\nu\, {\bar \mu}{\bar \nu}}$ is the retarded 
Green's function regularized according to the
Detweiler-Whiting prescription \cite{Detweiler:2002mi,introEMRI}.
The forced motion of the particle is then equivalent at linear order
to geodesic motion in the metric $g_{\mu\nu} + h_{R\ \mu\nu}$, where $Q'$
is held fixed when evaluating the geodesic equation and then evaluated
at $Q'=Q$ \cite{Detweiler:2002mi,pound}.
We can therefore obtain a pseudo-Hamiltonian description of the
dynamics by replacing the metric $g_{\mu\nu}(x)$ in the Hamiltonian
with $g_{\mu\nu}(x) + h_{R\ \mu\nu}(x,Q')$ and expanding to linear
order.  We can also specialize to including just the conservative
piece of the self-force, by replacing the regularized retarded Green's function
$G_R^{\mu\nu\,{\bar\mu}{\bar\nu}}$ with the average
$G^{\mu\nu\,{\bar\mu}{\bar\nu}}$ 
of the retarded and advanced Green's functions, regularized in the
same way, and replacing the metric perturbation $h_{R\ \mu\nu}$
with its conservative piece $h_{\mu\nu}$.

Turn now to the corresponding story for spinning point particles.
For the zeroth order motion we use phase space coordinates
$Q^A = (x^\mu, \pi_\mu,\sigma, \rho_\sigma, \varsigma, \rho_\varsigma)$ on $\Gamma$
defined in Eq.\ (\ref{eq:canonicalangles}),
with symplectic form (\ref{eq:canpoisson})
and Hamiltonian (\ref{eq:mpdham}).
This motion is described by the equations of motion (\ref{eq:mpd}) and is zeroth
order in self-gravity, but contains effects first order in spin.

For the first order motion, consider a particle at location $x^{\mu'}$
with initial 4-momentum $p_{\mu'}$ and initial spin
$S^{\Lambda'\Sigma'}$ [here the spin variable $S^{\Lambda'\Sigma'}$ should be understood
to be a shorthand for the four variables $\sigma, \rho_\sigma,
\varsigma, \rho_\varsigma$ defined in Eq.\ (\ref{eq:canonicalangles})].
Writing $Q' = (x^{\mu'},\pi_{\mu'},S^{\Lambda'\Sigma'})$, we denote
by
$
\varphi_{\tau'}(Q') = [ x^{{\bar \mu}}(\tau'), \pi_{{\bar
      \mu}}(\tau'),
S^{{\bar \Lambda}{\bar \Sigma}}(\tau')]
$
the solution to the zeroth order motion and spin evolution
(\ref{eq:mpd}), where $\tau'$ is proper time.
We can compute from this zeroth order motion a metric perturbation as
follows.  Inserting the stress energy tensor of a spinning point
particle given by Eq.\ (9) of Ref.\ \cite{Mathews:2021rod} into the
linearized Einstein equation gives the Lorenz gauge metric perturbation
\begin{eqnarray}
  \label{mp1}
  h^{\mu\nu}_R(x,Q')
  &=& 
\int d\tau'
G_R^{\mu\nu\, {\bar \mu}{\bar \nu}}[x, {\bar x}(\tau')]
\frac{p_{{\bar \mu}}(\tau') p_{{\bar \nu}}(\tau')}{\sqrt{ - g^{{\bar
        \lambda}{\bar \sigma}} p_{\bar\lambda} p_{\bar\sigma}}}
\nonumber \\ && 
-\int d\tau'
\nabla_{{\bar \rho}}G_R^{\mu\nu\, {\bar \mu}{\bar \nu}}[x, {\bar x}(\tau')]
\frac{p_{{\bar \mu}}(\tau') e_{{\bar \Lambda} {\bar \nu}} e_{\bar \Sigma}^{\ {\bar
      \rho}} S^{{\bar \Lambda}{\bar \Sigma}} }{\sqrt{ - g^{{\bar
        \lambda}{\bar \sigma}} p_{\bar\lambda} p_{\bar\sigma}}}.
\end{eqnarray}
Here barred indices indicate quantities that are evaluated at
$x^{\bar{\mu}}(\tau')$, and $\nabla_{\bar \rho}$ acts only on the
second argument of the Green's function.
The factor of $\sqrt{-{\vec p}^2}$ could be evaluated either at
$x^\mu$ or at $x^{\bar \mu}$, since it is conserved by the dynamics
(\ref{eq:mpd}); we choose the latter for later convenience.
Now it is known that the self-forced and self-torqued motion of the
spinning particle is given at linear order by evaluating the
equations of motion (\ref{eq:mpd}) 
in the metric $g_{\mu\nu} + h_{R\ \mu\nu}$, where $Q'$
is held fixed when evaluating the equations and then evaluated
at $Q'=Q$ \cite{Mathews:2021rod}.  It follows that 
we can obtain a pseudo-Hamiltonian description of the
dynamics by
making the replacements
\begin{subequations}
  \label{replace1}
  \begin{eqnarray}
    g_{\mu\nu}(x) &\to& g_{\mu\nu}(x) + h_{R\ \mu\nu}(x,Q') , \\
    e_\Lambda^{\ \mu} &\to& e_\Lambda^{\ \mu} - \frac{1}{2}
    e_\Lambda^{\ \nu} \, \, g^{\sigma\mu}\, h_{R\ \nu\sigma}(x,Q') 
  \end{eqnarray}
\end{subequations}
in the Hamiltonian (\ref{eq:mpdham})
    and expanding to linear
order.  Here the perturbation to the orthonormal basis is chosen to
maintain orthonormality.  Note that in order to apply the result of
Sec.\ \ref{sec:pseudoham}, we must use the form (\ref{eq:mpdham}), (\ref{eq:canpoisson}) of the dynamical
system for which the symplectic form is constant and so not modified by the
substitutions (\ref{replace1}), rather than the original form (\ref{eq:poisson}),
(\ref{eq:geoham}).  This is because
the result requires that the symplectic form be unperturbed,
cf.\ Eq.\ (\ref{php00}).

To complete the pseudo-Hamiltonian formulation of the dynamics,
we need to write the pseudo-Hamiltonian in terms of the phase space
variables $(x^\mu, \pi_\nu, S_{\Lambda\Sigma})$.
We start by writing the metric perturbation (\ref{mp1}) in terms of
the new momentum variable (\ref{pidef}) and expanding to linear order
in spin, which gives
\be
\begin{aligned}\label{eq:metricperturbation}
h_R^{\alpha\beta}(x,Q') &=h^{\alpha\beta}_{R(m)}(x,Q')+h^{\alpha\beta}_{R(S)}(x,Q'),
\end{aligned}
\ee
where
\begin{subequations}
  \label{eq:mp1}
  \begin{eqnarray}
\label{eq:metricperturbation1}
h^{\alpha\beta}_{R(m)}(x,Q')&=\int d\tau' G^{\alpha\beta\bar{\mu}\bar{\nu}}\big[x,\bar{x}(\tau')\big]\frac{\pi_{\bar{\mu}}(\tau')\pi_{\bar{\nu}}(\tau')}{\sqrt{-g^{\bar{\rho}\bar{\sigma}}\pi_{\bar{\rho}}\pi_{\bar{\sigma}}}},\\
h^{\alpha\beta}_{R(S)}(x,Q')&=\int d\tau'
G^{\alpha\beta\bar{\mu}\bar{\nu}}\big[x,\bar{x}(\tau')\big]\frac{
  \pi_{\bar{\mu}}\omega_{\bar{\nu}\bar{\Theta} \bar{\Pi}}
  S^{\bar{\Theta}\bar{\Pi}}}{\sqrt{-g^{\bar{\rho}\bar{\sigma}}\pi_{\bar{\rho}}\pi_{\bar{\sigma}}}}\nonumber
\\
&-\int d\tau' \nabla_{\bar \rho}
G^{\alpha\beta\bar{\mu}\bar{\nu}}\big[x,\bar{x}(\tau')\big]\frac{
  \pi_{\bar{\mu}}(\tau')
  e_{\bar{\Theta}\bar{\nu}}
  e_{\bar{\Pi}}^{\ \bar{\rho}}
  S^{\bar{\Theta}
    \bar{\Pi}}(\tau')}{\sqrt{-g^{\bar{\rho}\bar{\sigma}}\pi_{\bar{\rho}}\pi_{\bar{\sigma}}}}\nonumber
\\
&+\frac{1}{2}\int d\tau' G^{\alpha \beta \bar{\mu}\bar{\nu}}\big[x,\bar{x}(\tau')\big] \frac{\pi_{\bar{\mu}}\pi_{\bar{\nu}}\pi^{\bar{\rho}}\omega_{\bar{\rho}\bar{\Lambda}\bar{\Theta}}S^{\bar{\Lambda}\bar{\Theta}}}{\big[-g^{\bar{\sigma}\bar{\lambda}} \pi_{\bar{\sigma}} \pi_{\bar{\lambda}}\big]^{3/2}}.
\end{eqnarray}
\end{subequations}
Below we will need the metric perturbation $h_{R}^{\mu\nu}$ accurate
to $O(m)$ and $O(S)$, and we can neglect $O(m^2)$, $O(m S)$ and
$O(S^2)$ contributions.  Hence in the expressions (\ref{eq:mp1}) it is
sufficient to use the geodesic worldline rather than the solution to
Eqs.\ (\ref{eq:mpd}) which incorporates $O(S)$ corrections to the worldline. 
We next make the replacements (\ref{replace1}) in the Hamiltonian
(\ref{eq:mpdham}).  This yields the pseudo-Hamiltonian
\be\label{eq:pseudoham}
\begin{aligned}
\mathcal{H}^R(Q,Q')&=-\sqrt{-g^{\mu\nu}\pi_\mu \pi_\nu}+\frac{g^{\mu\nu}\pi_\mu \omega_{\nu \Theta \Pi}S^{\Theta \Pi}}{2\sqrt{-g^{\mu\nu}\pi_\mu \pi_\nu}}-\frac{h_R^{\mu\nu}(x,Q')\pi_\mu \pi_\nu }{2\sqrt{-g^{\mu\nu}\pi_\mu \pi_\nu }}\\
&-\frac{h_R^{\mu\nu}(x,Q')\pi_\mu \omega_{\nu \Theta \Pi}S^{\Theta \Pi}}{2\sqrt{-g^{\mu\nu}\pi_\mu \pi_\nu}}+\frac{g^{\mu\nu}\pi_\mu e_\Theta ^\alpha e_\Pi ^\beta h_{\nu[\alpha;\beta]}^R S^{\Theta \Pi}}{2\sqrt{-g^{\mu\nu}\pi_\mu \pi_\nu }}\\
&-\frac{1}{4}\frac{h_R^{\mu\nu}(x,Q')\pi_\mu\pi_\nu\pi^\rho\omega_{\rho\Lambda \Theta}S^{\Lambda\Theta}}{\big[-g^{\sigma\lambda} \pi_{\sigma} \pi_{\lambda}\big]^{3/2}},
\end{aligned}
\ee
where we used that the perturbation to the spin connection is $\delta \omega_{\mu \Lambda \Pi}=e_\Lambda^\alpha e_\Pi^\beta h_{\mu[\alpha;\beta]}$.

As an aside, we can verify as follows that the 
the pseudo-Hamiltonian (\ref{eq:pseudoham}) with symplectic form
(\ref{eq:canpoisson}) gives the correct dynamics 
for a spinning particle under the effect of the first order
gravitational self-force.
Using Eq.\ (\ref{phdef}) we
obtain for the equations of motion
$\nabla_{\vec{u}}u^\mu = a^\mu$ and 
$\nabla_{\vec{u}}S^{\mu\nu} = N^{\mu\nu}$,
where the self-acceleration $a^\mu$ and self-torque $N^{\mu\nu}$ are
given by
\begin{subequations}\label{eq:self-forcempd}
\begin{eqnarray}
a^\mu
&=&-\frac{1}{2}\big[g^{\mu\lambda}+u^{\mu}u^\lambda\big]\big[2h^R_{\lambda\rho;\sigma}-h^R_{\rho\sigma;\lambda}\big]u^\rho
u^\sigma \nonumber \\
      \nonumber  &-&\frac{1}{2m}R^{\mu}_{\ \alpha\beta\gamma}\big[1-\frac{1}{2}h^{R(m)}_{\rho\gamma}u^\rho u^\gamma\big]u^\alpha S^{\beta\gamma}\\
      &+&\frac{1}{2m}\big[g^{\mu\nu}+u^{\mu}u^\nu\big]\big[2h^{R(m)}_{\nu (\alpha;\beta)\gamma}-h^{R(m)}_{\alpha\beta;\nu\gamma}\big]u^\alpha S^{\beta\gamma},\\
        \label{eq:self-forcempdspin}N^{\mu\nu}&=&u^{(\rho}S^{\sigma)[\mu}g^{\mu]\lambda}\big[2h^{R(m)}_{\lambda\rho;\sigma}-h^{R(m)}_{\rho\sigma;\lambda}\big],
\end{eqnarray}
\end{subequations}
and where the metric perturbation $h^R_{\mu\nu}$ has been evaluated at
$Q'=Q$ after the derivatives have been taken.  These equations agree
with those of Ref.\ \cite{Mathews:2021rod}.
They can also be obtained by making the substitutions (\ref{replace1}) in the
equations of motion (\ref{eq:mpd}). 
As discussed in the introduction, we keep only terms of order $O(m^2)$
, $O(S)$ and $O(m S)$ in the self-force, and $O(m S)$ in the self-torque,
which explains why we have replaced $h^{\mu\nu}_R$ with
$h^{\mu\nu}_{R\,(m)}$ [cf.\ Eq.\ (\ref{eq:metricperturbation1})] in some of the terms in (\ref{eq:self-forcempd}).

\end{section}

\begin{section}{\label{sec:ham} Hamiltonian formulation of the
    conservative motion of a self-gravitating spinning particle}

In this section we
show that the motion of a spinning point particle
under the action of the first order conservative self-force is
Hamiltonian, by combining
the pseudo-Hamiltonian formulation of the motion derived in
Sec.\ \ref{sec:pseudohamforslefforce} with
the general result of Sec.\ \ref{sec:pseudoham}.
To do this we need to read off the function
$G(Q,Q')$ on phase space defined by Eqs.\ (\ref{php1}) and
(\ref{tildeGdef}), and to verify that it satisfies the required  
properties (\ref{condts}).

We start by specializing to the conservative sector of the dynamics.
As described after Eq.\ (\ref{hpp}) above in the nonspinning case, this
is achieved by replacing in the pseudo-Hamiltonian (\ref{eq:pseudoham})
the regularized retarded Green's function
$G_R^{\mu\nu\,{\bar\mu}{\bar\nu}}$ with the average
$G^{\mu\nu\,{\bar\mu}{\bar\nu}}$ 
of the retarded and advanced Green's functions, regularized in the
same way, and replacing the metric perturbation $h_{R\ \mu\nu}$
with its conservative piece $h_{\mu\nu}$.  Note that this Green's
function obeys the symmetry property
\begin{subequations}
  \begin{eqnarray}
\label{Gsym0}
    G^{\mu\nu\alpha'\beta'}(x,x')=G^{\alpha'\beta'\mu\nu}(x',x).
  \end{eqnarray}
\end{subequations}

Next, by comparing the pseudo-Hamiltonian given by Eqs.\ (\ref{eq:metricperturbation}) and
(\ref{eq:pseudoham}) with the general form given by Eqs.\ (\ref{php}),
(\ref{php1}) and (\ref{tildeGdef}), we obtain for the function
$G(Q,Q')$ on phase space
\begin{eqnarray}
  \label{G2}
      G(Q,Q')&=& \frac{1}{4} N N' \bigg[
      -2 \pi_\mu \pi_\nu \pi_{\rho'} \pi_{\sigma'} G^{\mu\nu\rho'\sigma'}(x,x')\nonumber \\
    &&-2 \pi_\mu \pi_\nu S^{\Theta' \Pi'}\pi_{\rho'}\omega_{\sigma'\Theta' \Pi'}
    G^{\mu\nu\rho'\sigma'}(x,x')\nonumber \\
    &&-2 \pi_{\rho'} \pi_{\sigma'}  S^{\Theta \Pi}\pi_{\mu}\omega_{\nu \Theta \Pi}
    G^{\mu\nu\rho'\sigma'}(x,x')\nonumber \\
    &&+2 \pi_\mu \pi_\nu \pi_{\rho'}e_{\Theta'\sigma'}e_{\Pi'}^{\ \lambda'}S^{\Theta' \Pi'}
    \nabla_{\lambda'} G^{\mu\nu\rho'\sigma'}(x,x')\nonumber \\
    &&+2 \pi_{\rho'} \pi_{\sigma'} \pi_{\mu}e_{\Theta\nu}e_{\Pi}^{\ \lambda}S^{\Theta \Pi}
      \nabla_{\lambda} G^{\mu\nu\rho'\sigma'}(x,x')\nonumber \\
      &&-N^{\prime\,2} \pi_\mu \pi_\nu \pi_{\alpha'}\pi_{\beta'}\pi_{\rho'}\omega^{\rho'}_{\ \Theta'\Pi'}S^{\Theta'\Pi'}
      G^{\mu\nu\alpha'\beta'}(x,x')\nonumber \\
      &&-N^2 \pi_{\alpha'} \pi_{\beta'}
      \pi_{\mu}\pi_{\nu}\pi_{\rho}\omega^{\rho}_{\ \Theta\Pi}S^{\Theta\Pi}
      G^{\mu\nu\alpha'\beta'}(x,x') \bigg],
\end{eqnarray}
where
\be
N = \frac{1}{{\sqrt{-g^{\alpha\beta}\pi_\alpha\pi_\beta}}}, \ \ \ \ \ N' = \frac{1}{{\sqrt{-g^{\rho'\sigma'}\pi_{\rho'} \pi_{\sigma'}}}}.
  \ee
Because of the symmetry property (\ref{Gsym0}) of the Green's function, the
function (\ref{G2}) satisfies the required symmetry property (\ref{condtA}).  It
also satisfies the required asymptotic conditions (\ref{condtB}) for the
reasons discussed in Ref.\ \cite{Blanco:2022mgd}.

It now follows from the result reviewed in section \ref{sec:pseudoham}
that the dynamical
system (\ref{eq:self-forcempd}) admits a
Hamiltonian description.  The Hamiltonian function is
given by Eqs.\ (\ref{eq:hamresult0}), (\ref{eq:hamresult}),
(\ref{eq:mpdham}), (\ref{tildeGdef}) and
(\ref{G2}), and the symplectic form is given by
(\ref{eq:canpoisson}), in phase space coordinates given by Eq.\ 
(\ref{eq:diffeo}).

\end{section}

\begin{section}{\label{sec:conclusions} Conclusions}

We have shown that the conservative dynamics of the two body problem
in general relativity is Hamiltonian to the first sub-leading order in
the mass and spin of the secondary. 
This result may yield useful information for the effective one body framework and thereby aid waveform modeling for comparable mass binary systems for LIGO. Extending this result to the second-order $O(m^2)$ point-particle 
self-force is a direction that is currently being explored. 

\end{section}

\vspace{0.1cm}
\noindent \textit{Acknowledgments:} We thank Vojt\v{e}ch Witzany
and Paul Ramond for helpful
discussions and references.
This
research was supported in part by NSF grant PHY-2110463.

\color{black}

\bibliography{Ref.bib}

\end{document}